\documentclass[showpacs,aps,pra,floats,twocolumn]{revtex4-1}
 \pdfoutput=1
\usepackage{graphicx}
\usepackage{braket}
\usepackage{color}
\usepackage{epstopdf}
\usepackage{mathtools}
\usepackage{amsmath}
\usepackage{amssymb}
\usepackage{hyperref}
\usepackage{float}
\graphicspath{{./figs/}}
\usepackage{gensymb}
\usepackage[colorinlistoftodos,prependcaption]{todonotes}
\usepackage{xargs}  
\newcommandx{\unsure}[2][1=]{\todo[linecolor=red,backgroundcolor=red!25,bordercolor=red,#1]{#2}}
\newcommandx{\change}[2][1=]{\todo[linecolor=blue,backgroundcolor=blue!25,bordercolor=blue,#1]{#2}}
\newcommandx{\info}[2][1=]{\todo[linecolor=OliveGreen,backgroundcolor=OliveGreen!25,bordercolor=OliveGreen,#1]{#2}}
\newcommandx{\improvement}[2][1=]{\todo[linecolor=Plum,backgroundcolor=Plum!25,bordercolor=Plum,#1]{#2}}
\newcommandx{\thiswillnotshow}[2][1=]{\todo[disable,#1]{#2}}
\newcommandx{\greencom}[2][1=]
{\todo[inline, color=green!40,#1]{#2}}
\newcommandx{\bluecom}[2][1=]
{\todo[inline, color=blue!40,#1]{#2}}

\begin{document}
% change equation skips - too big in align default
\abovedisplayskip=7pt
\abovedisplayshortskip=0pt
\belowdisplayskip=7pt
\belowdisplayshortskip=7pt

\newcommand{\sigp}{\sigma^+}
\newcommand{\sigm}{\sigma^-}
\newcommand{\Gspph}{\Gamma^{\sigma^+}_0}
\newcommand{\Gsmph}{\Gamma^{\sigma^-}_0}
\newcommand{\Gcdph}{\Gamma^\mathrm{cd}_0}
\newcommand{\vc}[1]{{\boldsymbol{\mathrm{#1}}}}
\newcommand{\comment}[1]{}
\newcommand{\remove}[1]{}
\newcommand{\quot}[1]{\textquotedblleft{}#1\textquotedblright}
\newcommand{\un}{\mathrm}
\newcommand{\B}{\langle B \rangle}
\newcommand{\blue}[1]{{\color{blue}#1}}
\newcommand{\red}[1]{{\color{red}#1}}
\makeatletter
\newcommand{\vast}{\bBigg@{4}}
\newcommand{\Vast}{\bBigg@{5}}
\newcommand{\G}{\mathbf{G}}
\makeatother
\title{Pulsed excitation dynamics in quantum dot-cavity systems: limits to optimizing the fidelity of on-demand single photon sources}
\author{Chris Gustin}
\email{c.gustin@queensu.ca}
\author{Stephen Hughes}
\affiliation{\hspace{-40pt}Department of Physics, Engineering Physics, and Astronomy, Queen's University, Kingston, Ontario K7L 3N6, Canada\hspace{-40pt}}
\date{\today}

\begin{abstract}
A quantum dot coupled to an optical cavity has recently proven to be an excellent source of on-demand single photons. Typically, applications require simultaneous high efficiency of the source and quantum indistinguishability of the extracted photons. While much progress has been made both in suppressing background sources of decoherence and utilizing cavity-quantum electrodynamics to overcome fundamental limitations set by the intrinsic exciton-phonon scattering inherent in the solid-state platform, the role of the excitation pulse has been often neglected. We investigate quantitatively the factors associated with pulsed excitation that can limit simultaneous efficiency and indistinguishability, including excitation of multiple excitons, multi-photons, and pump-induced dephasing, and find for realistic single photon sources that these effects cause degradation of the source figures-of-merit comparable to that of phonon scattering. We also develop rigorous open quantum system polaron master equation models of quantum dot dynamics under a time-dependent drive which incorporate non-Markovian effects of both photon and phonon reservoirs, and explicitly show how coupling to a high Q-factor cavity suppresses multi-photon emission in a way not predicted by commonly employed models. We then use our findings to summarize the criteria that can be used for single photon source optimization.
\end{abstract}
\maketitle

\section{Introduction}\label{sec0}

Recently, the use of photons as easily manipulated and decoherence-resistant quanta for quantum information processing applications, as well as fundamental studies, has led  to the desire to create a high-fidelity source which can produce single photons on demand~\cite{senellart17,kuhn10,muller17,shields07,broome13}. The single photon source should produce photons efficiently (ideally, where each trigger pulse produces one and only one photon) which are of an indistinguishable quantum character (i.e., a pure quantum state), in addition to other desirable parameters including scalability and source stability over time. Of much success in providing a physical realization of a single photon source in recent years is the semiconductor quantum dot (QD)~\cite{buckley12,ding16,somaschi16,he17,liu17} -- an effectively zero-dimensional semiconductor material, where the electronic bandgap and three-dimensional confinement allows for excited electron-hole pairs (excitons) that behave like the ground and an excited state of an artificial atom. Exploiting this optically-active transition, an optical pulse can create an exciton which then radiatively relaxes to the ground state, emitting a single photon. The solid-state QD single photon source has the advantage of being stable, as well as easily integrated in photonic environments (e.g. optical cavities), but comes with additional challenges as decoherence sources unique to the solid-state degrade the indistinguishability and efficiency of emitted photons~\cite{lodahl15}. Decoherence reduces the purity of the quantum state by coupling to a large number of degrees of freedom; this can include charge and spin noise from the semiconductor material~\cite{kuhlmann13}, which recent experimental techniques have been shown to successfully suppress~\cite{somaschi16,kuhlmann15}, and most notably, electron-phonon coupling intrinsic to the dot~\cite{nazir16}. Intrinsic electron-phonon scattering fundamentally alters the field-dipole interaction by introducing a reservoir of phonon modes (most significantly, longitudinal acoustic (LA) modes) which couple to the two-level system, introducing non-Markovian decoherence, including a broad phonon sideband in the emission spectrum, and decoherence rates which scale with the pulse intensity in the excitation process~\cite{ramsay10,krummheuer02,besombes01,vagov02,forstner03,ross16}. Phonon-induced decoherence thus leads to limits on the simultaneous efficiency and indistinguishability of the QD as a single photon source, which can be partially mitigated by coupling the QD to an optical cavity, accelerating the emission of photons into the zero-phonon-line (ZPL) via Purcell enhancement, simultaneously increasing the collection efficiency and indistinguishability in frequency of the emitted single photons~\cite{ilessmith17,grange17}. However, the phonon interaction means that these figures-of-merit can not be simultaneously increased without limit. This limit to simultaneous efficiency and indistinguishability has been recently investigated in the case of an initially inverted QD~\cite{ilessmith17}. The role of the excitation pulse required in real single-photon sources to initially excite the QD, however, has been relatively neglected. In particular, the probability of two-photon emission becomes non-zero in the presence of a finite pulse excitation~\cite{fischer17,2fischer17}, which degrades the indistinguishability of the emitted wavepacket, and excitation-induced dephasing rates caused by phonon scattering degrade the single photon efficiency. Given the interest in optimizing such sources and understanding intrinsic limits, this neglect is quite questionable and is clearly of interest to the community.

In this work, we demonstrate  explicitly how the pulse parameters can interact with the exciton, the cavity mode, and the phonon bath to degrade the single photon source figures-of-merit, and in particular, how the source's environment and the pulse can be engineered to minimize these drawbacks. Combining these findings with  previous work on how cavity-quantum electrodynamics (QED) can be used to minimize phonon-induced decoherence, we report criteria for pulse and cavity parameters to optimize simultaneous maximization of efficiency and indistinguishability for QD-cavity single photon sources. Additionally, we present different methods of modelling time-dependent QD dynamics with a master equation approach, rigorously incorporating cavity (and other photonic reservoirs) and phonon coupling. While pulse-driven systems have been previously modelled with master equations~\cite{mccutcheon10,ross16,gustin17,gustin18,moelbjerg12} incorporating cavity and phonon coupling, the extent to which the pulse can induce non-Markovian decoherence has typically been neglected, and as such excitation regimes with pulse durations shorter than the correlation times of the system-reservoir interactions have been less accessible. We investigate the extent to which these additional Markov assumptions regarding the driven system influence the dynamics and then apply this to the excitation process of QD single photon sources.

The layout of the paper is as follows: in Sec.~\ref{sec1}, we introduce the theoretical model of our system via a cavity-QED polaron master equation technique and clarify the definition of the quantum indistinguishability of single photons in terms of Hong-Ou-Mandel interference; in Sec.~\ref{sec2}, we review the phonon-limits to simultaneous efficiency and indistinguishability and investigate the role of the excitation pulse on the single-photon figures-of-merit by analyzing the effects of multi-photon emission and multiple exciton excitation; in Sec.~\ref{sec3} we show how the presence of a cavity mode suppresses multi-photon emission in a way not predicted by the traditional ``bad-cavity" approximation, and explain this effect by developing a non-Markovian master equation approach which self-consistently captures the effect of a short pulse on a QD interacting with arbitrary weakly-coupled photonic and phononic environments. In Sec.~\ref{sec4}, we calculate the numerical limits to simultaneous efficiency and indistinguishability obtainable when the pulse excitation process is considered, and summarize criteria regarding the cavity and pulse parameters to aid in optimal single photon source design. We then conclude in Sec.~\ref{sec5}. We also include two appendices where we give a full cavity-scattering term for our master equation without an approximation made in the main text, as well as a simplified phonon-scattering term valid for weak phonon coupling.

\section{Cavity-QED polaron master equation}\label{sec1}

\begin{figure}[ht]
\centering
\includegraphics[width=1\linewidth]{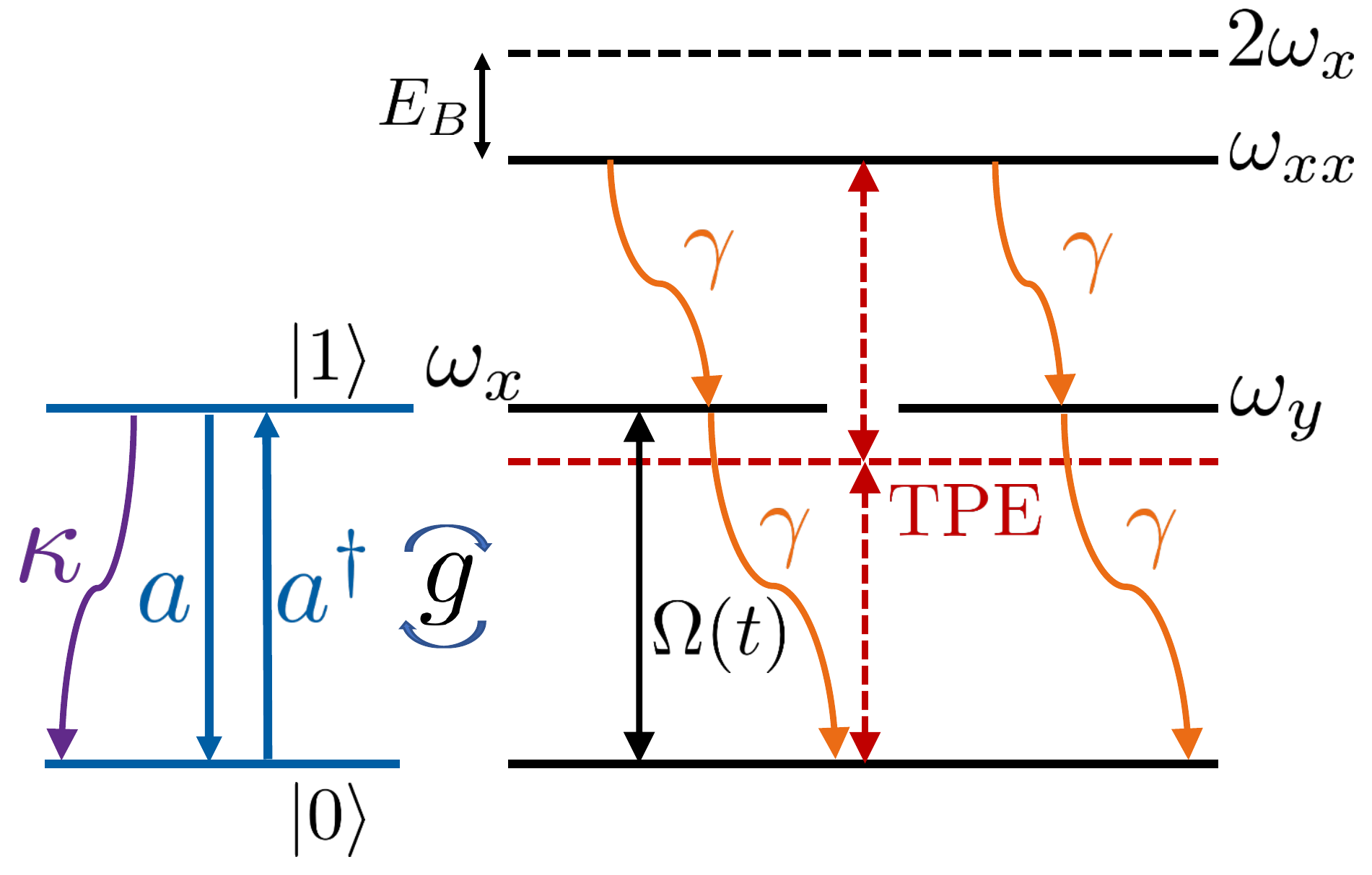}
\caption{\small Schematic of a four-level cavity-QED model with the biexciton-exciton cascade. The two-photon excitation (TPE) resonance condition is detuned $E_B/2$  below the exciton energy level. The biexciton cascade system contains radiative decay channels from the biexciton-to-exciton, exciton-to-ground, as well as from the biexciton to an orthogonally linearly polarized exciton (denoted with $y$), and from this $y$-exciton to ground; the units use  $\hbar =1$. Both excitons and the biexciton are coupled to a bath of LA phonon modes (not shown), and for simplicity we only show the first two states of the cavity mode operator.}
\label{fig0} 
\end{figure}

In this section we introduce a model of the cavity-QD system that uses a four-level polaron master equation model of exciton and biexciton energy states coupled to a quantized cavity mode, which we will refer to as the cavity-QED model for the sake of differentiating our later approach (where the cavity is treated as a bath). Exciton-phonon scattering is conveniently modelled with a polaron master equation approach, where the zero-field exciton-phonon coupling is incorporated nonperturbatively via the independent Boson model~\cite{mahan} by unitary transformation into a ``polaron" frame, and the pump (or cavity)-polaron scattering is then treated with usual Born-Markov master equation techniques~\cite{mccutcheon11,roy11,roy12}. This approach has several advantages over alternative theoretical approaches, including  weak coupling master equations~\cite{nazir16}, which break down for strong phonon coupling rates and elevated temperatures, numerically exact path integral methods~\cite{barth16,dattani13}, which are less well developed for calculating two-time correlation functions, and variational master equations~\cite{mccutcheon11}, which may break down for general pulse shapes and short pulse lengths. As we will show below, short pulses are required to suppress multi-photon emission from the QD-cavity system. However, a lower bound on the pulse lengths is placed by the presence of the biexciton (two exciton) state~\cite{kiraz04}, that has a binding energy $E_B = 2\hbar\omega_{x}-\hbar\omega_{xx}$ ($x$ for exciton, $xx$ for biexciton) which in typical QDs is small ($\sim\text{meV}$ scale) enough that short pulses have large enough spectral bandwidth to excite the two-photon resonance condition ($E_B/2$ detuned below the exciton) to excite the biexciton state (see Fig.~\ref{fig0}). To incorporate this effect, we model in this section our QD-cavity system as a four-level biexciton cascade scheme (where the additional state is an orthogonally polarized ($y$) exciton, populated via spontaneous emission from the biexciton), coupled to a quantized cavity mode with creation (destruction) operator $a^{\dagger}$ ($a$). Since the biexciton can decay to either polarization of exciton, then to ground, this reduces the number of the cavity emitted photons, as well as the indistinguishability via a timing jitter. 

Throughout this work, we
exclusively consider the case where the pulse carrier frequency is resonant with the $x$-exciton, such that the rotating-frame total Hamiltonian (neglecting for now spontaneous emission and cavity photon loss) $H$ of the QD + cavity + phonons system is
\begin{align}\label{eq1}
H =& -E_B\ket{xx}\bra{xx} + \frac{\hbar\Omega(t)}{2}(\sigma_x + \ket{xx}\bra{x} + \ket{x}\bra{xx})  \nonumber \\ &+ \hbar g(\sigma^{+}a + \sigma^{-}a^{\dagger} + \ket{xx}\bra{x}a + \ket{x}\bra{xx}a^{\dagger})  \nonumber \\  &+ \sum\limits_{\mathbf{q}}\hbar\omega_{\mathbf{q}}b_{\mathbf{q}}^{\dagger}b_{\mathbf{q}}+ \ \sum\limits_{\mathclap{s=x,y,xx}} \ \ket{s}\bra{s}\sum\limits_{\mathbf{q}}\hbar\lambda_{\mathbf{q}}^{s}(b_{\mathbf{q}}^{\dagger}+b_{\mathbf{q}}),
\end{align}
in terms of the ground-exciton pseudospin Pauli operators ($\sigma_x = \sigma^+ + \sigma^-$, $\sigma_y = i(\sigma^--\sigma^+)$, $\sigma_z = \sigma^+\sigma^- - \sigma^-\sigma^+$), and bosonic operators $b$, $b^\dagger$ corresponding to phonon modes with wavevector $\mathbf{q}$. The LA phonon-exciton coupling is included via coupling constants $\lambda_{\mathbf{q}}^{s}$ (assumed real)  for $s = \{x,y,xx\}$, which  correspond to an ideal quantum confined QD such that $\lambda_{\mathbf{q}} \equiv \lambda_{\mathbf{q}}^{x} = \lambda_{\mathbf{q}}^{y} = \frac{1}{2}\lambda_{\mathbf{q}}^{xx}$~\cite{hohenester07}. To treat electron-phonon coupling nonperturbatively, we carry out a unitary polaron transformation $H' = e^P H e^{-P}$, with~$P =  (\sigma^+ \sigma^- + \ket{y}\bra{y} + 2\ket{xx}\bra{xx}) \sum\limits_{\mathbf{q}} \frac{\lambda_{\mathbf{q}}}{\omega_{\mathbf{q}}}(b^\dagger_{\mathbf{q}} - b_{\mathbf{q}})$ to diagonalize the electron-phonon coupling part of the Hamiltonian, and derive a perturbative master equation to deal with pulse and cavity-induced fluctuations around this new polaron system~\cite{mahan,mccutcheon10}. We can separate the new polaron transformed Hamiltonian into system, bath, and coupling parts such that $H' = H_S' + H_B' + H_I'$, where the RWA system Hamiltonian becomes
\begin{align}
H_S' = & -E_B\ket{xx}\bra{xx} + \frac{\hbar\Omega'(t)}{2}(\sigma_x + \ket{xx}\bra{x} + \ket{x}\bra{xx})  \nonumber \\ &+ \hbar g'(\sigma^{+}a + \sigma^{-}a^{\dagger} + \ket{xx}\bra{x}a + \ket{x}\bra{xx}a^{\dagger}),
\end{align}
where we have ignored a small polaron shift in the exciton and biexciton energies, assuming it to be absorbed into the definition of $E_B$. The cavity coupling and Rabi drive become renormalized by the coherent displacement average of the phonon bath such that $g' = \B g$, $\Omega'(t) = \B \Omega(t)$, with  $\B = \langle B_+ \rangle = \langle B_- \rangle 
= \text{exp}\Big[-\frac{1}{2}\sum_{\mathbf{q}}\frac{\lambda^2_{\mathbf{q}}}{\omega^2_{\mathbf{q}}}\coth{\big(\frac{\hbar\omega_{\mathbf{q}}}{2k_B T}\Big)}\Big]$, at temperature $T$, as the expectation value of the displacement operators $ B_{\pm} = \text{exp}\big[\pm\sum_{\mathbf{q}}\frac{\lambda_{\mathbf{q}}}{\omega_{\mathbf{q}}}(b_{\mathbf{q}}^{\dagger}-b_{\mathbf{q}})\big]$. The bath Hamiltonian is $H_B' = \sum\limits_{\mathbf{q}}\hbar\omega_{\mathbf{q}}b_{\mathbf{q}}^{\dagger}b_{\mathbf{q}}$, and the interaction Hamiltonian becomes $H_I' = X_g\zeta_g + X_u \zeta_u$, with $X_g = \frac{\hbar\Omega(t)}{2}(\sigma_x + \ket{xx}\bra{x} + \ket{x}\bra{xx}) + \hbar g (\sigma^{+}a + \sigma^{-}a^{\dagger} + \ket{xx}\bra{x}a + \ket{x}\bra{xx}a^{\dagger})$, $X_u = i\frac{\hbar\Omega(t)}{2}(i\sigma_y + \ket{xx}\bra{x} - \ket{x}\bra{xx}) + i\hbar g(\sigma^+ a - \sigma^- a^\dagger + \ket{xx}\bra{x}a - \ket{x}\bra{xx}a^{\dagger})$, and phonon fluctuation operators $\zeta_g = \frac{1}{2}(B_+ + B_- - 2\B)$, $\zeta_u = \frac{1}{2i}(B_+ - B_-)$. 

In the continuum limit of phonon modes, we can characterize the electron-phonon interaction with the phonon spectral function $J_p(\omega) = \sum_{\mathbf{q}}\lambda_{\mathbf{q}}^2\delta(\omega-\omega_{\mathbf{q}}) \rightarrow J_p(\omega) = \alpha \omega^3 \text{exp}\big[\frac{\omega^2}{2\omega_b^2}\big]$, which describes a deformation potential induced by LA phonons---the main source of phonon-related decoherence in solid-state QDs such as GaAs and InAs~\cite{nazir16,ramsay10,2ramsay10}; $\alpha$ is the exciton-phonon coupling strength, and $\omega_b$ is the phonon cut-off frequency. A polaron master equation is derived in this frame:
\begin{align}\label{me}
\frac{d}{dt}\rho(t) &= -\frac{i}{\hbar}[H'_S, \rho(t)] - \frac{1}{\hbar^2}\int_0^{\infty}d\tau \sum\limits_{\mathclap{m = g, u}}  \big(G_m(\tau) \times \nonumber  \\
&\!\!\!\!\!\!\!\!\!\!\!\!\!\! [X_m(t), \widetilde{X}_m(t-\tau,t) \rho(t)] + {\rm H.c.} \big) + \frac{1}{2}\sum_{\mu} \mathcal{L}[O_\mu]\rho(t),
\end{align}
which has been previously used to analyze pulse-driven phonon-exciton scattering in biexciton systems~\cite{gustin17}. Here, $G_g(\tau) = \B^2 (\cosh{(\phi(\tau))}-1)$ and $G_u(\tau) = \B^2\sinh{(\phi(\tau))}$ are the polaron Green functions, and $\phi(\tau) = \int\limits_{0}^{\infty}d\omega \frac{J_p(\omega)}{\omega^2}\left(\coth{\left(\frac{ \hbar \omega}{2 k_B T}\right)}\cos{(\omega\tau)} - i\sin{(\omega\tau)}\right)$. 

The interaction picture operators $ \widetilde{X}_m(t-\tau,t)$ are typically approximated $ \widetilde{X}_m(t-\tau,t)\approx e^{-iH'_S(t)\tau/\hbar} X_m(t) e^{iH'_S(t)\tau/\hbar}$, which we shall refer to as an ``additional Markov approximation,'' although in Sec.~\ref{sec4} we calculate them exactly. Note that the upper limit of the integral has been extended to infinity; while this is often considered a ``second Markov approximation", equivalent to the assumption that correlations in the system-environment interaction decay on a shorter timescale than the system dynamics~\cite{breuer}, in our case this is simply the correct initial condition for our setup. Often, it is assumed that the system has been prepared in the excited state at $t =0$, under which circumstances the upper limit of the integral becomes $t$ directly from integrating the Von-Neumann equation, which can then be extended to $\infty$ via a second Markov approximation. In our case, we have incorporated the excitation process directly into the model, and thus we extend the limit to $\infty$ to insert the condition that the system is in the ground state at $t\rightarrow -\infty$. We also include phenomenological Lindblad dissipation terms (for collapse operator $O$: $\mathcal{L}[O]\rho = 2O\rho O^{\dagger} - O^\dagger O \rho - \rho O^\dagger O$) with collapse operators $\sqrt{\gamma}\ket{x}\bra{xx}$, $\sqrt{\gamma}\ket{y}\bra{xx}$, $\sqrt{\gamma}\ket{g}\bra{x}$, $\sqrt{\gamma}\ket{g}\bra{y}$, and $\sqrt{\kappa}a$ corresponding to spontaneous emission and cavity photon leakage. Initially the (electrically-neutral) QD is taken to be in the ground state and the cavity mode to be in the vacuum state. 

\subsection{Figures-of-Merit for Single Photon Sources}

The two main figures-of-merit of interest to this study are the {\it efficiency} of the single photon source and {\it indistinguishability} of the cavity-emitted photons. The quantitative metric of efficiency studied here is really the emitted cavity photon number -- in terms of the cavity mode operators, this value is 
\begin{equation}
N_c = \int_{0}^{\infty}\kappa \langle a^\dagger a \rangle (t) dt.
\end{equation}
Note this is only the cavity emission efficiency; the total single-photon efficiency of course involves efficiencies in coupling to outgoing modes. Also of interest is the so-called $\beta$-factor, which can be calculated as 
\begin{equation}
\beta = \frac{N_c}{N_c+N_x},
\end{equation}
where 
\begin{equation}
N_x = \int_{0}^{\infty}\gamma \langle \sigma^+ \sigma^- \rangle (t) dt,
\end{equation}
is the exciton-emitted photon number (photons emitted into non-cavity modes). For a high efficiency source, in principle, only $N_c$ needs to be as close to unity as possible, although in practice $\beta$ must also be close to $1$ to avoid drops in single-photon purity via multi-photon emission. Single-photon purity (vanishing probability of multi-photon emission), as is measured in a Hanbury-Brown-Twiss interferometery setup, is also an important criterion for single-photon sources, but our definition of indistinguishability encapsulates this requirement.

\subsection{Single photon Indistinguishability}
The single-photon indistinguishability is a measure of the purity of the quantum state of the emitted photon. For a source where the probability of emitting more than one photon is zero (e.g., a QD which is prepared in the excited state and allowed to radiatively decay), this is a measure of the first-order coherence of the source; the spectrum of each emitted photon wavepacket is identical to the previous one. For pulse-triggered sources, the multiphoton probability is typically non-zero, and the indistinguishability of the quantum state is also a function of the second-order (intensity) coherence. To provide an experimentally-accessible metric of indistinguishability, the phenomena of two-photon interference is typically probed via a Hong-Ou-Mandel (HOM) interferometry setup. Here, two photons emitted from identical single-photon sources are incident upon a beam splitter, and the cross-correlation function of photodetectors placed at the output channels is measured. For a single-photon source which emits indistinguishable photons in their first and second order coherences, the cross-correlation function at zero delay vanishes. In practice, a Mach-Zehnder interferometer can be used to simulate the HOM setup with only a single photon source, triggered with a delay much longer than the lifetime of the exciton~\cite{2fischer16}. In this case, we can model the input channels to the beamsplitter in terms of field operators proportional to the cavity-mode operators $a$, $a^\dagger$. Note that these operators are unaffected by the polaron transform, so the indistinguishability can be calculated from them directly. The first and second order coherences are modelled via the two-time correlation functions $g^{(1)}(t,\tau)=\langle a^{\dagger}(t)a(t+\tau)\rangle$ and $g^{(2)}(t,\tau)= \langle a^{\dagger}(t)a^{\dagger}(t+\tau)a(t+\tau)a(t)\rangle$, respectively. The intensity cross-correlation of the output channels $G^{(2)}_{\text{HOM}}(t,\tau)$ is then~\cite{kiraz04,woolley13,2fischer16}
\begin{equation}
G^{(2)}_{\text{HOM}}(t,\tau) = \frac{1}{2}\big(G_{\text{pop}}^{(2)}(t,\tau) + g^{(2)}(t,\tau) - |g^{(1)}(t,\tau)|^2\big),
\end{equation}
where $G_{\text{pop}}^{(2)}(t,\tau)=\langle a^{\dagger}a\rangle(t)\langle a^{\dagger}a\rangle(t+\tau)$. Consider a single-photon source triggered with period $2T$, where $T$ is long enough that the single photon source has returned to its ground state. Since $G^{(2)}_{\text{HOM}}(t,\tau)$ goes to zero around $\tau = 0$ for a perfect single-photon source, it makes sense to define an indistinguishability $\mathcal{I}$ (or two-photon interference visibility) as follows:
\begin{equation}
\mathcal{I} = 1 - \frac{\int_0^T dt \int_{-T}^T d\tau  G^{(2)}_{\text{HOM}}(t,\tau)}{\int_0^T dt \int_{T}^{3T} d\tau   G^{(2)}_{\text{HOM}}(t,\tau)}.
\end{equation}
If the cross-correlation is time-averaged over $t$, then this corresponds to taking the ratio of the area on the plot of $G^2(\tau)$ of the peak around $\tau = 0$ to the peak around $\tau = 2T$ and subtracting it from unity. This is the typical experimental procedure (up to corrections due to, e.g., beamsplitter imperfections)~\cite{somaschi16}. To calculate $\mathcal{I}$ with only a single pulse excitation, note that for $\tau > T$, $g^{(2)}(t,\tau)$ and $g^{(1)}(t,\tau)$ turn into a product of uncorrelated expectation values $ \langle a^{\dagger}a\rangle(t)\langle a^{\dagger}a\rangle(t+\tau)= G_{\text{pop}}^{(2)}(t,\tau)$ and $\langle a(t+\tau)\rangle \langle a^{\dagger}(t)\rangle$, respectively, such that
\begin{equation}\label{three}
\mathcal{I} =1 - \frac{\int_0^T dt \int_{0}^T d\tau  \big(G_{\text{pop}}^{(2)}(t,\tau) + g^{(2)}(t,\tau) - |g^{(1)}(t,\tau)|^2\big)}{\int_0^T dt \int_{0}^{T} d\tau   \big(2G_{\text{pop}}^{(2)}(t,\tau) - |\langle a(t+\tau)\rangle \langle a^{\dagger}(t)\rangle|^2\big)},
\end{equation}
where negative values of $\tau$ have been excluded due to the symmetry of the peaks, and we have used the fact that the one-time expectation values are periodic with period $2T$. Often the second term in the denominator is neglected, allowing for further simplification, as it is small for a good single-photon source where the photon emission does not significantly occur during the excitation process. In the limit of a perfectly incoherent single-photon source (large pure dephasing), that is in the excited state at time $t = 0$, this definition of indistinguishability tends to $1/2$, and so a different definition $\mathcal{I'} = 2\mathcal{I} -1$ is sometimes used~\cite{ilessmith17}. For our calculations, we employ Eq.~\eqref{three}, evaluating the two-time correlation functions with the aid of the quantum regression theorem~\cite{carmichael}.

\section{Limits to single photon efficiency and indistinguishability}\label{sec2}

\subsection{Influence of phonon scattering}\label{phononlimits}
Previous works have thoroughly studied the detrimental effects of phonon-coupling on single-photon figures-of-merit for an initially inverted QD and shown how cavity-QED can be used to partially circumvent this source of decoherence~\cite{ilessmith17,grange17}. The most notable effect of exciton-phonon coupling in QDs is to create a broad phonon sideband in the emission spectrum, with spectral width on the order of the phonon cutoff frequency $\omega_b$. In contrast to the Lorentzian ZPL of a two-level system emitting radiatively into a spectral reservoir, the phonon sideband arises from incoherent and real phonon transitions during the photon emission process (dominated at low temperatures by phonon-emission-assisted radiative decay). As a result, indistinguishable photons must be extracted from the ZPL. A frequency-based post-selection around the ZPL accomplishes this, but with a decay in efficiency which reduces the on-demand nature of the single-photon source. By instead introducing coupling with a high Purcell factor cavity, one can accelerate the emission of photons into the ZPL relative to the free-space decay rate, thus simultaneously increasing both the efficiency of the source (via the $\beta$-factor), and the indistinguishability of the emitted photons. Restrictions are placed on the extent to which this effect can be harnessed by cavity parameters; the Purcell factor, usually equal to $F_P = \frac{4g'^2}{\kappa \gamma}$, can be increased by decreasing the cavity decay rate $\kappa$ by introducing higher quality-factor cavities, or increasing the QD-cavity coupling strength $g$. However, the decay rate must be much smaller than the phonon cutoff frequency to ensure effective filtering of the phonon sideband ($\kappa \ll \omega_b$), and the ratio $4g'/\kappa$ should be kept not too large to minimize phonon-induced dephasing in the cavity-QD interaction~\cite{ilessmith17}. Lastly, a larger decay rate minimizes exposure times to both phonon and non-phonon dephasing processes.

To show some of the conclusions above more explicitly, neglecting the biexciton state, we can adiabatically eliminate the cavity mode from our above master equation, assuming that we are dealing with the dynamics long after the pulse has decayed to zero, and we are in the weak-coupling regime $4g' \ll \kappa$~\cite{ilessmith17}. The cavity operators can then be approximated by letting $\frac{\rm{d}a}{\rm{dt}} \approx 0$ in the Heisenberg equation of motion, yielding $a \approx -2i\frac{g'}{\kappa}\sigma^-$. For $g' \ll \omega_b$, we can also approximate the transformed operators $\widetilde{X}_m(t-\tau,t) \approx X_m(t)$. Substituting this into Eq.~\eqref{me}, we find
\begin{equation}
\frac{\rm{d}\rho}{\rm{dt}} = \frac{\gamma}{2}(1+F_P)\mathcal{L}[\sigma^-]\rho + \frac{\gamma_c'}{2}\mathcal{L}[\sigma^+\sigma^-]\rho -i[\Delta_c\sigma^+\sigma^-,\rho],
\end{equation}
with $F_P = \frac{4g'^2}{\kappa\gamma}$, $\gamma_c' = 2\text{Re}\{\Lambda_c\}$, $\Delta_c = \text{Im}\{\Lambda_c\}$, where 
\begin{equation}
\Lambda_c = \big(\frac{4g'^2}{\kappa}\big)^2\int_0^{\infty}d\tau\sinh{(\phi(\tau))}.
\end{equation}
It is thus clear that in the limit of weak coupling with the cavity, the phonon interaction induces a pure dephasing rate and Lamb-type shift that both scale with $(\frac{g'^2}{\kappa})^2$.

\subsection{Influence of the excitation pulse}
In addition to the phonon limits discussed in section \ref{phononlimits}, the presence of the optical pulse required to excite an on-demand QD single-photon source also places {\it often-neglected restrictions} on the simultaneous maximization of efficiency and indistinguishability. Most significantly, if the decay rate of the excited system is comparable to the temporal length of the excitation pulse, the probability of emitting a multi-photon wavepacket from a single excitation pulse becomes significant, degrading the purity of the single-photon source and thus the HOM indistinguishability via the second-order coherence function. Radiative emission during the excitation process furthermore degrades the first-order coherence of the source via dephasing. Recent work has shown the probability of two-photon emission in spectrally-flat reservoirs to scale with the product of the pulse width $\tau_p$ and radiative decay rate $\gamma$ for small two-photon probabilities~\cite{fischer17}, and Eq.~\eqref{three} suggests the same scaling with respect to the indistinguishability of the emitted wavepacket. The effective decay rate into the cavity mode should be large, as to minimize phonon-induced decoherence, so the pulse width should thus be as small as possible to avoid emission during the excitation process. A lower bound, however, is placed on this pulse width by the presence of 
higher-lying energy states in the QDs -- namely multi-exciton states. 

\begin{figure}[!t]
\centering
\includegraphics[width=1\linewidth]{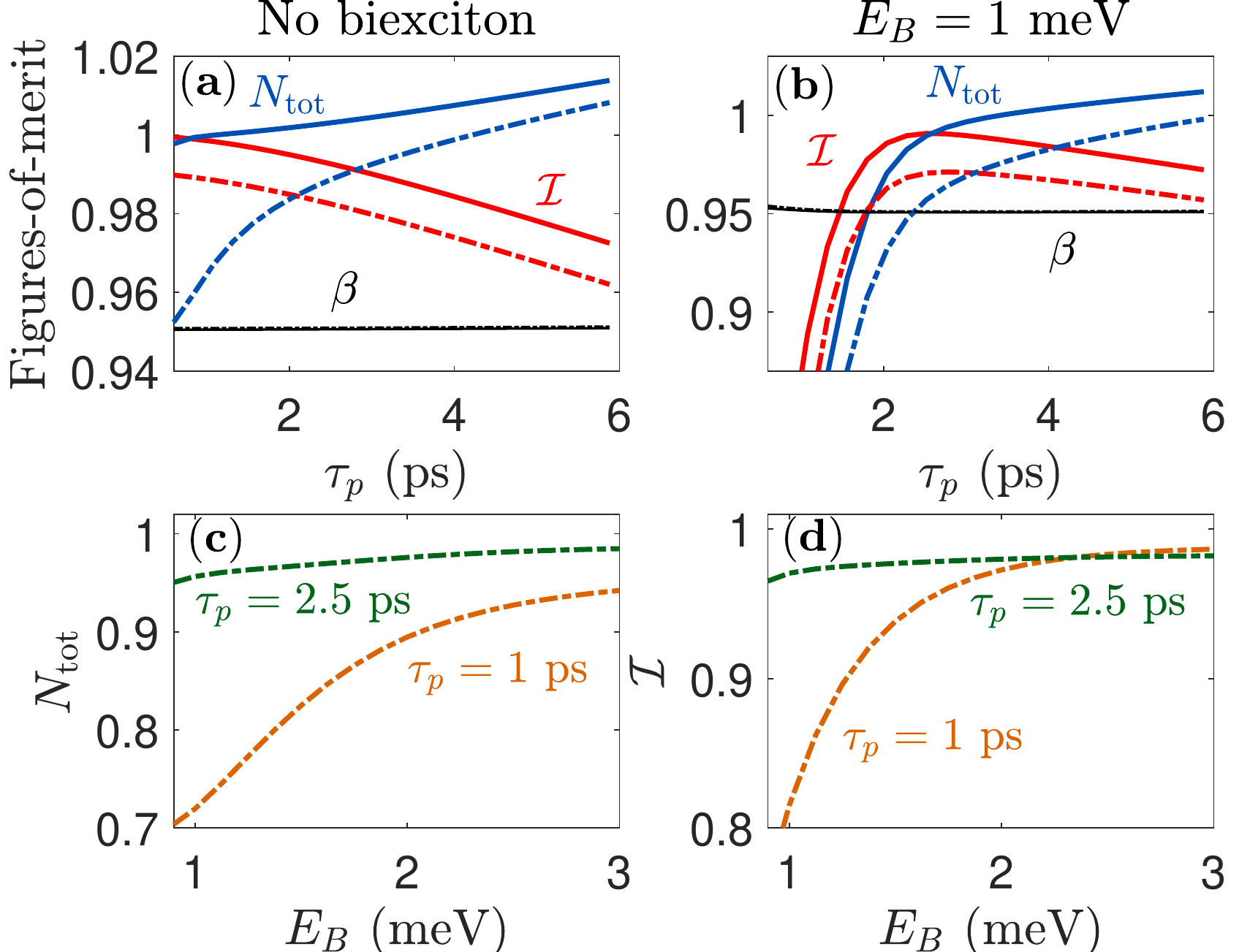}
\caption{\small Top: emitted photon number and indistinguishability as a function of pulse width for (a) a two-level system model and (b) a four-level system with biexciton binding energy $E_B = 1 \ \text{meV}$, both without phonons (solid line) and with (dash-dotted). Bottom: total emitted photon number (c) and indistinguishability (d) for two different pulse durations as a function of biexciton binding energy, both with phonons. Note that the sudden decrease in emitted photon number for pulses with $\tau_p \lesssim 2 \ \text{ps}$ with phonons is mostly an artifact of the additional Markov approximation. Here $\hbar g' = 25 \ \mu \text{eV}$ and $\hbar \kappa = 250 \ \mu \text{eV}$, such that the Purcell factor $F_P = \frac{4g'^2}{\kappa \gamma} = 20$} 
\label{fig1} 
\end{figure}

We included the biexciton state in our cavity-QED model in Sec.~\ref{sec1}, expecting the pulses to lower the efficiency and indistinguishability of the source when the spectral width of the pulse becomes comparable to the binding energy $E_B$. To test our expectations, we plot in Fig.~\ref{fig1} the indistinguishability $\mathcal{I}$, the total number of emitted photons from the QD, $N_{\text{tot}} = N_c + N_x$, and the $\beta$-factor $\beta = N_c/N_{\text{tot}}$. Throughout this work, we employ a Gaussian pulse shape
\begin{equation}
\Omega(t) = \frac{\Theta}{\sqrt{\pi} \tau_p} e^{-\frac{(t-3\tau_p)^2}{\tau_p^2}}, 
\end{equation}
where $\Theta = \int_{-\infty}^{\infty} dt\, \Omega(t)$ is the pulse area, and $\tau_{\text{FWHM}}=2\sqrt{\text{ln}(2)}\tau_p$. We use the phonon parameters $\alpha = 0.03 \ \text{ps}^2$, and $\hbar\omega_b = 0.9 \ \text{meV}$, similar to those found from the experimental
results by Quilter \emph{et al.}~\cite{quilter15}, and the temperature $T = 4 \ \text{K}$. The free-space decay rate is $\hbar\gamma = 0.5 \ \mu\text{eV}$~\cite{hargart16}, and the phonon-renormalized pulse is a pi-pulse (i.e., $\Theta  =\pi/\B$). As expected, Fig~\ref{fig1} reveals that increasing pulse durations lead to a nearly linear increase in the multi-photon probability (which can be deduced from the behavior of $N_{\text{tot}}$) and a decrease in indistinguishability. Phonon-coupling decreases the overall emitted photon number and indistinguishability -- the former effect can be attributed to pump-induced dephasing effects during the excitation process, while the latter can be attributed to analogous cavity-induced dephasing. The $\beta$-factor is nearly unaffected by phonon-coupling, indicating that phonons do not play a significant role in the radiation dynamics into various photon modes -- an observation confirmed by our analysis in Sect.~\ref{sec3}  -- and is very nearly equal to its analytical value obtained from an initially inverted quantum dot~\cite{cui06}: 
\begin{equation}
\beta = \frac{F_P}{F_P+1}\,\frac{1}{1+\gamma/\kappa},
\end{equation}
with $F_P = 4g'^2/(\kappa \gamma)$. The sudden drop in emitted photon number with the phonon interaction for short pulses is mostly an unphysical artifact of our {\it additional Markov approximation} (see Fig.~\ref{fig4}). We also investigate here the effect of different biexciton binding energies, as this can vary significantly from QD-to-QD. The regime in which the probability of two-photon excitation of the biexciton is negligible can be determined by requiring that the frequency component of the Fourier-transformed pulse at the two-photon resonance condition is much smaller than the exciton resonance: in our rotating frame, ${\Omega(\omega=\frac{E_B}{2\hbar})}/{\Omega(\omega=0)}= \exp{\big[-(\frac{\tau_p E_B}{4\hbar})^2\big]} \ll 1$, or alternatively, $\frac{\tau_p E_B}{4\hbar} \gg 1$. 

As increasing the pulse width leads to a decreasing indistinguishability due to the linearly increasing two-photon emission probability, one might wonder if the same decrease is seen with an increasing Purcell factor. Critically, this is \emph{not} the case. In fact, if the increase in the Purcell factor is caused by decreasing the cavity decay rate $\kappa$, the two-photon emission probability can decrease. Importantly, this effect is usually only seen with the cavity mode treated at the system level with operators $a, a^\dagger$ -- the usual ``bad-cavity" approximation in which the cavity mode is adiabatically eliminated simply replaces the two-level system decay rate $\gamma$ with $(F_P+1) \gamma$ (although adiabatically eliminating the cavity does retain other terms corresponding to electron-phonon scattering~\cite{ilessmith17}), and thus {\it incorrectly} predicts an increase in the two-photon emission probability that scales approximately linearly with $F_P \gamma \tau_p$. This is due to the time-dependent nature of the pulsed excitation. To explain further, we develop in the following section a self-consistent system-reservoir theory, and demonstrate how this approach can be used to derive more accurate master equations for resonant excitation which capture important non-Markovian effects in the interaction with arbitrary photon and phonon reservoirs.

\section{System-Reservoir Theory Approach}\label{sec3}

In this section, we drop the $a, a^\dagger$ quantized cavity operators (treated at the system level), and instead consider a two-level system (neglecting now the potential for biexciton excitation) interacting with photon and phonon reservoirs. Using this formal system-reservoir model, we can derive a time-dependent master equation which shows explicitly the effect of phonon-coupling and the excitation pulse on the radiation dynamics of the QD-cavity system. This reveals how the cavity interaction allows for the Purcell factor to be increased almost arbitrarily without increasing the probability of multi-photon emission, and thus, not causing the corresponding reduction in single-photon purity and indistinguishability. Furthermore, we expect the approach we take here to be useful in going beyond standard polaron master equations which have to date assumed Markovianity in the pump-induced electron-phonon scattering, allowing for modelling of weakly-coupled systems (not just cavity systems) in which the system dynamics unfold on similar or smaller timescales than the reservoir dynamics. This approach also allows for an updated bad-cavity approximation which retains pump-induced effects. Note also that we are effectively extending the model of Ref.~\cite{2roychoudhury16} to allow for a time-dependent drive on resonance.

We begin with the total Hamiltonian of a two-level pulse-driven system interacting with a photon and phonon bath, rotating at the exciton frequency~\cite{roychoudhury15,2roychoudhury16,2roychoudhury15}:
\begin{align}
H = &\hbar\int d\mathbf{r} \int_{0}^{\infty} d\omega\, \omega \mathbf{f}^\dagger  (\mathbf{r},\omega) \!\cdot\! \mathbf{f}(\mathbf{r},\omega)  + \frac{\hbar \Omega(t)}{2} \sigma_x \nonumber \\ & - \Bigg[\sigma^+ e^{i\omega_x t} \int_{0}^{\infty} d \omega \mathbf{d} \!\cdot\! \mathbf{E}(\mathbf{r_d},\omega) + \text{H.c.}\Bigg] \nonumber \\ & + \sum\limits_{\mathbf{q}}\hbar\omega_{\mathbf{q}}b_{\mathbf{q}}^{\dagger}b_{\mathbf{q}} + \sigma^+\sigma^-\sum\limits_{\mathbf{q}}\hbar\lambda_{\mathbf{q}}(b_{\mathbf{q}}^{\dagger}+b_{\mathbf{q}}),
\end{align}
which is similar to Eq.~\eqref{eq1} but with the cavity-dot interaction terms instead replaced with a general interaction term containing the electric field operator $\mathbf{E}(\mathbf{r},\omega)$, and a bath of photonic modes continuous in frequency with bosonic operators $\mathbf{f}(\mathbf{r},\omega)$, $\mathbf{f}^\dagger(\mathbf{r},\omega)$. Here, the QD is considered within the dipole and rotating wave approximations to be located at position $\mathbf{r_d}$, and the electric field is expressed in terms of the photonic Green function for the cavity~\cite{dung98} $\G (\mathbf{r},\mathbf{r'};\omega)$  (although this can be adopted for other reservoirs as well, including photonic crystal waveguides~\cite{2roychoudhury15}), as well as free field terms: $\mathbf{E}(\mathbf{r_d},\omega) = \mathbf{E}_{\text{free}}(\mathbf{r_d},\omega)+ i\int d\mathbf{r'}\G(\mathbf{r_d},\mathbf{r'};\omega)\!\cdot\! \mathbf{f}(\mathbf{r'},\omega)  \sqrt{\frac{\hbar}{\pi \epsilon_0}\epsilon_I(\mathbf{r'},\omega)}$, where $\epsilon_I$ is the imaginary component of the medium's dielectric constant. 
%although we neglect the free terms going forward, as their effect is included phenomenologically via the spontaneous emission rate. \blue{The main point is that they will not affect the spectrum results or any results with the chosen ordering, but they could, even with the SE term. Another option is to say we do not separate out free terms as they play no role in what follows below - the SE is just added by hand, but would come in through another set of f operators; f here is for the structured/interesting reservoir, not including the background /boring one}. 
The polaron transform of Sect.~\ref{sec1} is again applied, and here we now separate the total Hamiltonian into system, reservoir (phonon + photon), cavity interaction, and phonon interaction parts such that $H' = H'_S + H'_R + H'_{c} + H'_{p}$. These are:
\begin{subequations}
\begin{align}
H'_S &= \frac{\hbar \Omega'(t)}{2}\sigma_x,\\
H'_R &= \hbar\int d\mathbf{r} \int_{0}^{\infty} d\omega \omega \mathbf{f}^\dagger  (\mathbf{r},\omega) \!\cdot\! \mathbf{f}(\mathbf{r},\omega) + \sum\limits_{\mathbf{q}}\hbar\omega_{\mathbf{q}}b_{\mathbf{q}}^{\dagger}b_{\mathbf{q}}, \\
H'_c &= -\Big[B_+\sigma^+ e^{i\omega_x t} \int_{0}^{\infty} d \omega 
\mathbf{d} \! \cdot \! \mathbf{E}(\mathbf{r_d},\omega) + \text{H.c.}\Big],\\
H'_{p} &= \sum\limits_{m=u,g} X_m (t) \zeta_m.
\end{align}
\end{subequations}
Here, $X_g(t) = \frac{\hbar\Omega(t)}{2}\sigma_x$, and $X_u(t) = -\frac{\hbar\Omega(t)}{2}\sigma_y$. We can now derive a master equation in this frame by tracing over both the photon and phonon reservoirs~\cite{roychoudhury15}. Denoting the interaction picture with tildes, we obtain the following time-convolutionless master equation for the reduced density operator $\tilde{\rho}$:
\begin{align}\label{eqnine}
\frac{\text{d}\tilde{\rho}(t)}{\text{dt}} = & -\frac{1}{\hbar^2}\int_0^\infty  d\tau \text{Tr}_{\text{c}}\text{Tr}_{\text{p}}\big([\tilde{H}'_{c} (t) + \tilde{H}'_{p} (t) , \nonumber \\ &  [\tilde{H}'_c (t-\tau)+\tilde{H}'_p (t-\tau),\tilde{\rho}(t)\rho_{\rm c}\rho_{\rm p}]]\big),
\end{align}
where we have again extended the integration limit to infinity to ensure the correct initial condition, and made a 2nd-order Born-Markov approximation by assuming the total interaction picture density matrix factorizes -- an assumption we can expect to be valid in the weak-coupling regime $4g' \ll \kappa$ in the photon reservoir coupling, and in the polaron regime $\frac{1}{2}(\frac{\Omega(t)}{\omega_b})^2(1-\langle B \rangle^4) \ll 1 $ in the phonon reservoir coupling~\cite{mccutcheon10}. 
Moreover, we assume the phonon reservoir $\rho_{\text{p}}$ to be a thermal state, and the cavity reservoir $\rho_{\text{c}}$ to be the vacuum state. As a consequence, we have the relations $\text{Tr}_{\text{p}}(\tilde{H}'_p \rho_{\text{p}})=\text{Tr}_{\text{c}}(\tilde{H}'_c \rho_{\text{c}})=0$. This allows us to split the master equation into a cavity-scattering and phonon-scattering part; transforming back to the Schr\"{o}dinger picture, we have
\begin{equation}
\frac{\text{d}\rho(t)}{\text{dt}} = -i[H'_S(t),\rho(t)] + \mathbb{L}_{c}\rho(t) + \mathbb{L}_{p}\rho(t) + \frac{\gamma}{2}\mathcal{L}[\sigma^-]\rho(t),
\end{equation}
where we have also added in a free-space (background) spontaneous emission decay term, and  $\mathbb{L}_c $ and $\mathbb{L}_p$ are superoperators corresponding to cavity and phonon scattering, respectively; note that both of these terms are in general affected by both the phonon and photon interaction, and contain incoherent scattering effects and small coherent renormalizations to the system Hamiltonian. 
\subsection{Exciton-phonon scattering}

The phonon-scattering term $\mathbb{L}_p \rho$ is similar to what is found in Sect.~\ref{sec1}, and can be found by tracing over the photon and phonon reservoirs and transforming back to the Schr\"{o}dinger picture, noting the time-dependence of the unitary operators involved:
\begin{align}
& \mathbb{L}_p\rho(t)  = \frac{1}{\hbar^2}\sum\limits_{m=u,g} \int_0^\infty d\tau \big(G_m(\tau) \times  \nonumber \\ &\big[\widetilde{X}_m(t-\tau,t) \rho(t)X_m(t)\! -\!X_m(t)\widetilde{X}_m(t-\tau,t)\rho(t)\big]\!+\! \text{H.c.}\big),
\end{align}
where $ \widetilde{X}_m(t-\tau,t) = U^\dagger(t-\tau,t)X_m(t-\tau)U(t-\tau,t)$. The unitary operator $U(t,t')$ evolves the state of a two-level system with time-dependent Hamiltonian $H'_S(t)$ from time $t'$ to time $t$. If we make the additional Markov approximations $\Omega(t-\tau) \approx \Omega(t)$, and $U(t-\tau,t) \approx  \text{exp}\big[\frac{i}{\hbar} H'_S(t)\tau\big]$, this recovers previous results~\cite{mccutcheon10}. However, as QD single-photons sources are typically driven with pulses with widths on the order of the phonon interaction timescale, this misses non-Markovian effects associated with the exciton-phonon reservoir interaction. Instead, we note that in the case of an on-resonant pulse, the system Hamiltonian commutes with itself at all times and the unitary operator can be solved analytically:
\begin{align}
U(t-\tau,t) &= \text{exp}\bigg[-iR(t,\tau)\sigma_x\bigg] \nonumber \\ & = \cos{\big[R(t,\tau)\big]}I - i\sin{\big[R(t,\tau)\big]}\sigma_x,
\end{align}
where $R(t,\tau) \equiv \frac{1}{2}\int_t^{t-\tau}\Omega'(t')dt'$ and $I$ is the identity operator. Analytic solutions are unavailable for the off-resonant pulse case, but this model can be extended by computing the transformed operators by solving the Heisenberg equation with Hamiltonian $H'_S(t)$ in the $\tau$ variable for each value of $t$, similar to the calculation of two-time correlation functions via the quantum regression theorem, which we do for the cavity-QED model in Sect.~\ref{sec4}. Proceeding with this result, we can simplify this phonon-scattering part of the master equation to a more insightful form:
\begin{align}
 \mathbb{L}_p\rho & = \frac{\gamma_p(t)}{2}\big[\mathcal{L}[\sigma^+]\rho + \mathcal{L}[\sigma^-]\rho\big] \nonumber \\ &+\zeta_p(t)\big[\sigma^+ \rho \sigma^+ + \sigma^- \rho \sigma^-\big] \nonumber \\ & +\big(i\Gamma^R_u(t)[\sigma^+\sigma^-\rho\sigma^- - \sigma^+\sigma^-\rho\sigma^+ - \sigma^-\rho] + \text{H.c.}\big) \nonumber \\  & +\big(\Gamma^I_u(t)[\sigma^+\sigma^-\rho\sigma^+ - \sigma^+\sigma^-\rho\sigma^- + \rho\sigma^-]+ \text{H.c.}\big),
\end{align}
where
\begin{equation}
\gamma_p(t)\! =\! \frac{\Omega(t)}{2}\!\!\!\int_0^\infty \!\!\! \!d\tau \Omega(t-\tau) \text{Re}\big\{G_g(\tau) + \cos{\big[2R(t,\tau)\big]}G_u(\tau)\big\},
\end{equation}

\begin{equation}
\zeta_p(t) \!= \!\frac{\Omega(t)}{2}\!\!\!\int_0^\infty \!\!\!\!d\tau \Omega(t-\tau) \text{Re}\big\{G_g(\tau) - \cos{\big[2R(t,\tau)\big]}G_u(\tau)\big\},
\end{equation}
and $\Gamma^R_u(t)$ and $\Gamma^I_u(t)$ are real rates such that $\Gamma_u = \Gamma^R_u + i\Gamma^I_u$, with

\begin{equation}
\Gamma_u(t) = -\frac{\Omega(t)}{2} \int_0^\infty d\tau \Omega(t-\tau)G_u(\tau)\sin{\big[2R(t,\tau)\big]}.
\end{equation}
In the case of weak phonon coupling (low temperatures and/or small phonon coupling constant) and the additional Markov approximation, $\mathbb{L}_p$ takes on a simpler form, described in Appendix~\ref{appA}.

\subsection{Exciton-cavity scattering}
We now move to the part of the master equation corresponding to the interaction of the exciton with the photonic reservoir $\mathbb{L}_c \rho$. From the cavity interaction terms in Eq.~\eqref{eqnine}, we trace over the photon and phonon reservoirs: 
\begin{align}
\mathbb{L}_{c}\rho = & \int_0^\infty d\tau \Big(A_R(\tau) \nonumber \\ & \times \big[\sigma^+\rho\sigma^-(t-\tau,t) - \sigma^+\sigma^-(t-\tau,t)\rho\big] + \text{H.c.}\Big),
\end{align}
with $\sigma^-(t-\tau,t) =U^\dagger(t-\tau,t) \sigma^-U(t-\tau,t)$. To arrive at this result, we have again assumed a vacuum state for the photonic reservoir, and used the canonical boson commutation relations $\mathbf{f}(\mathbf{r},\omega)\!\cdot\! \mathbf{f}^{\dagger}(\mathbf{r'},\omega') -  \mathbf{f}^{\dagger}(\mathbf{r'},\omega') \!\cdot\! \mathbf{f}(\mathbf{r},\omega)= \delta(\mathbf{r}-\mathbf{r'})\delta(\omega-\omega')$, as well as the identity $\int d\mathbf{s} \epsilon_I (\mathbf{s},\omega) \G(\mathbf{r},\mathbf{s};\omega)\!\cdot\! \G^* (\mathbf{s},\mathbf{r'};\omega)=\text{Im}\big\{ \G(\mathbf{r},\mathbf{r'};\omega)\big\}$~\cite{dung98}. The function $A_R(\tau) = J_c(\tau) C_p(\tau)$ is the product of the photon and phonon bath correlation functions, respectively, with $C_p(\tau) = \langle B \rangle^2 e^{\phi(\tau)}=\langle B_+ (\tau)B_-(0)\rangle$~\cite{mahan}, and $J_c(\tau) = \int_0^{\infty} d\omega J_c(\omega)e^{-i(\omega-\omega_x)\tau}$, where $J_c(\omega) = \frac{1}{\pi \epsilon_0 \hbar} \mathbf{d}\!\cdot\! \text{Im}\big\{ \G(\mathbf{r_d},\mathbf{r_d};\omega)\big\}\!\cdot\! \mathbf{d}$~\cite{ge13} (assuming ${\bf d}$ is real). For low temperatures ($T\sim 4$ K), $C_p(\tau)$ deviates little from its steady-state value $C_p(\infty) = \langle B \rangle^2$, so we can approximate $C_p(\tau) \approx \langle B \rangle ^2$; we give in Appendix~\ref{appB} the full result without this approximation. Defining $J'_c(\tau) \equiv \langle B \rangle^2 J_c(\tau)$ and assuming a symmetric photon spectral correlation function such that $J'_c(\tau)$ is real, we can again simplify:
\begin{align}
\mathbb{L}_c \rho& = \frac{\Gamma_c(t)}{2}\mathcal{L}[\sigma^-]\rho + \zeta_c(t)(\sigma^+\rho\sigma^+ + \sigma^- \rho \sigma^-) \nonumber \\ & + i\int_0^\infty d\tau J'_c(\tau)\sin{\big[2R(t,\tau)\big]} \nonumber \\ & \times (\sigma^+\sigma^-\rho\sigma^+ - \sigma^- \rho\sigma^+\sigma^-)  -i[\frac{\tilde{\Omega}(t)}{2}\sigma_x,\rho],
\end{align}
where 
\begin{equation}\label{gamma}
\Gamma_c(t) = 2 \int_0^{\infty} d\tau J'_c(\tau) \cos^2{\big[R(t,\tau)\big]},
\end{equation}

\begin{equation}
\zeta_c(t)= \int_0^{\infty} d\tau J'_c(\tau) \sin^2{\big[R(t,\tau)\big]},
\end{equation}
and
\begin{equation}
\tilde{\Omega}(t)= - \int_0^{\infty} d\tau J'_c(\tau) \sin{\big[2R(t,\tau)\big]}.
\end{equation}

We can then define the effective time-dependent Purcell factor as $F_p(t) = {\Gamma_c(t)}/{\gamma}$.

\begin{figure}[!t]
\centering
\includegraphics[width=1\linewidth]{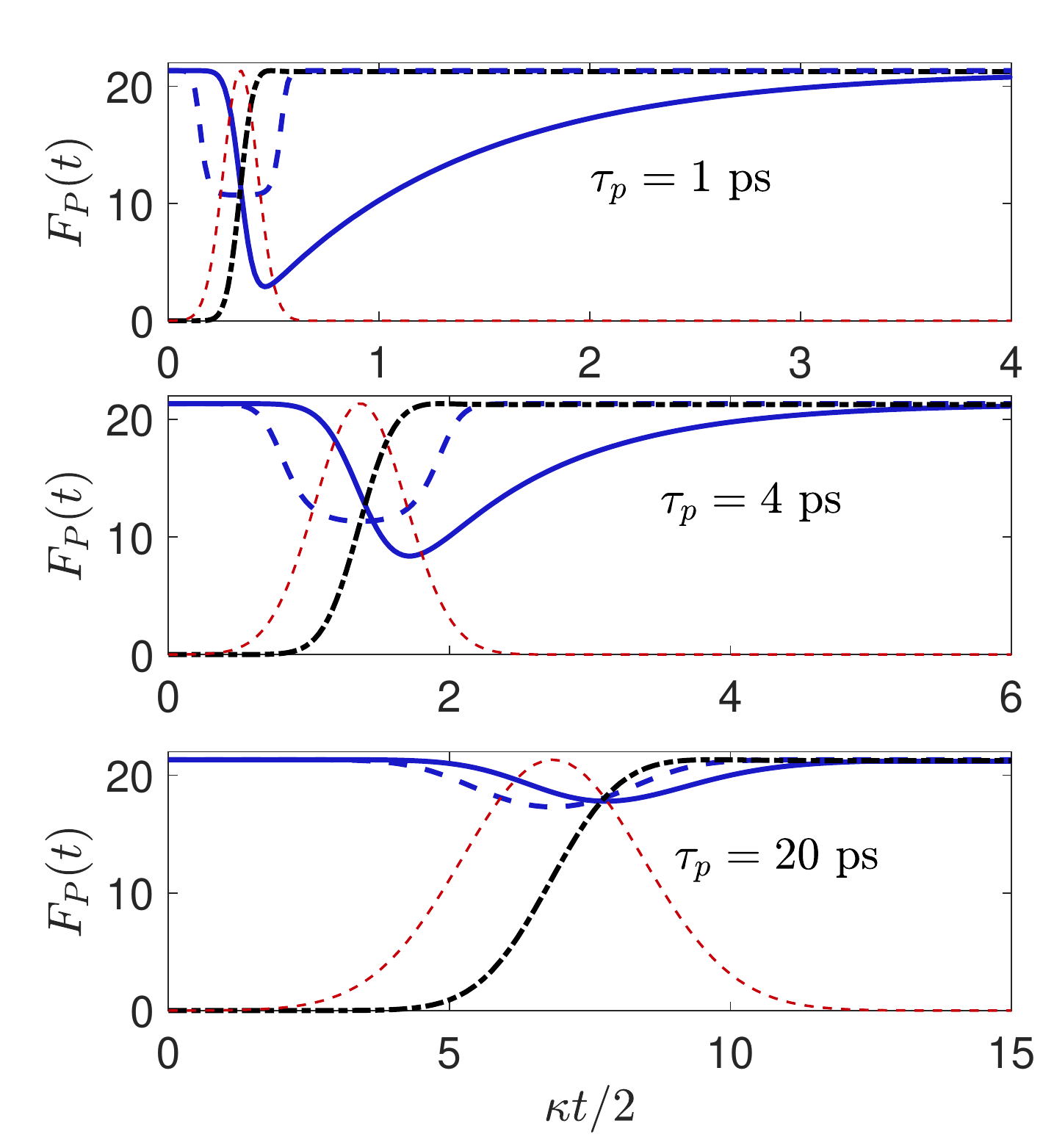}
\caption{\small Time-dependent Purcell factor (blue) for three different pulse widths, with the previously used additional Markov approximation (dashed), and with the full solution incorporating memory effects associated with the pulse (solid; Eq.~\eqref{gamma}). Also shown is the pulse envelope (red dashed), as well as the corresponding exciton population neglecting any dissipation, $\langle \sigma^+\sigma^-\rangle\propto\sin^2{\big(\frac{1}{2}\int_0^{t}\Omega'(t')dt'\big)}$ (black dash-dotted), both in arbitrary units. For these simulations, $\hbar g' = 20 \ \mu \text{eV}$ and $\hbar \kappa =150 \ \mu \text{eV}$.}
\label{fig2} 
\end{figure}

In this work we study the Lorentzian spectral function for a ``cavity'' bath function on resonance with the exciton~\cite{roychoudhury15}:
$J'_c(\tau) = g'^2e^{-\frac{\kappa}{2}\tau}$. If we consider the limit of a Delta-function pulse (neglecting excitation of other excitons) of the form $\Omega'(t)=\pi\delta(t-t_0)$, the {\it time-dependent Purcell factor} takes on the simple form 
\begin{equation}
F_P(t) = \frac{4g'^2}{\kappa \gamma}\left (1-e^{-\frac{\kappa}{2}(t-t_0)}\right ) ,
\end{equation}
for $t \geq t_0$, which in the long term limit recovers the familiar expression $F_P(\infty) = {4g'^2}/{\kappa \gamma}$. However, in contrast to equilibrium time-independent systems where this expression can be applied in the weak-coupling regime, for systems driven with short pulses, the pulse width $\tau_p$ can be much smaller than ${1}/{\kappa}$, and the time-dependent expression is essential. To illustrate this point, we plot in Fig.~\ref{fig2} the time-dependent Purcell factor for different pulse widths, revealing the suppression of spontaneous emission into the cavity during the pulse for $\kappa \tau_p \ll 1$ ($1/e$ full pulse width much less than $\frac{2}{\kappa}$). The previously used additional Markov approximations fail to capture the extent of this effect, as well as the non-Markovian build up of the spontaneous emission rate once the pulse has decayed, and the temporal location of the maxima/minima of the scattering rates. The analytic solution for the exciton population in the absence of any dissipation is also shown to help illustrate temporally the magnitude of the suppression, as the total emission rate into the cavity is the product of the exciton population and $\Gamma_c(t)$. In the long pulse limit, the full result recovers that of the additional Markov approximation. Note that for larger pulse areas, the differences between the full result and the additional Markov approximation become more pronounced; in the full result, $\Gamma_c(t)$ exhibits Rabi oscillations following the exciton dynamics, whereas with the additional Markov approximation it effectively averages these out~\cite{ross16}.

\begin{figure}[!t]
\centering
\includegraphics[width=1\linewidth]{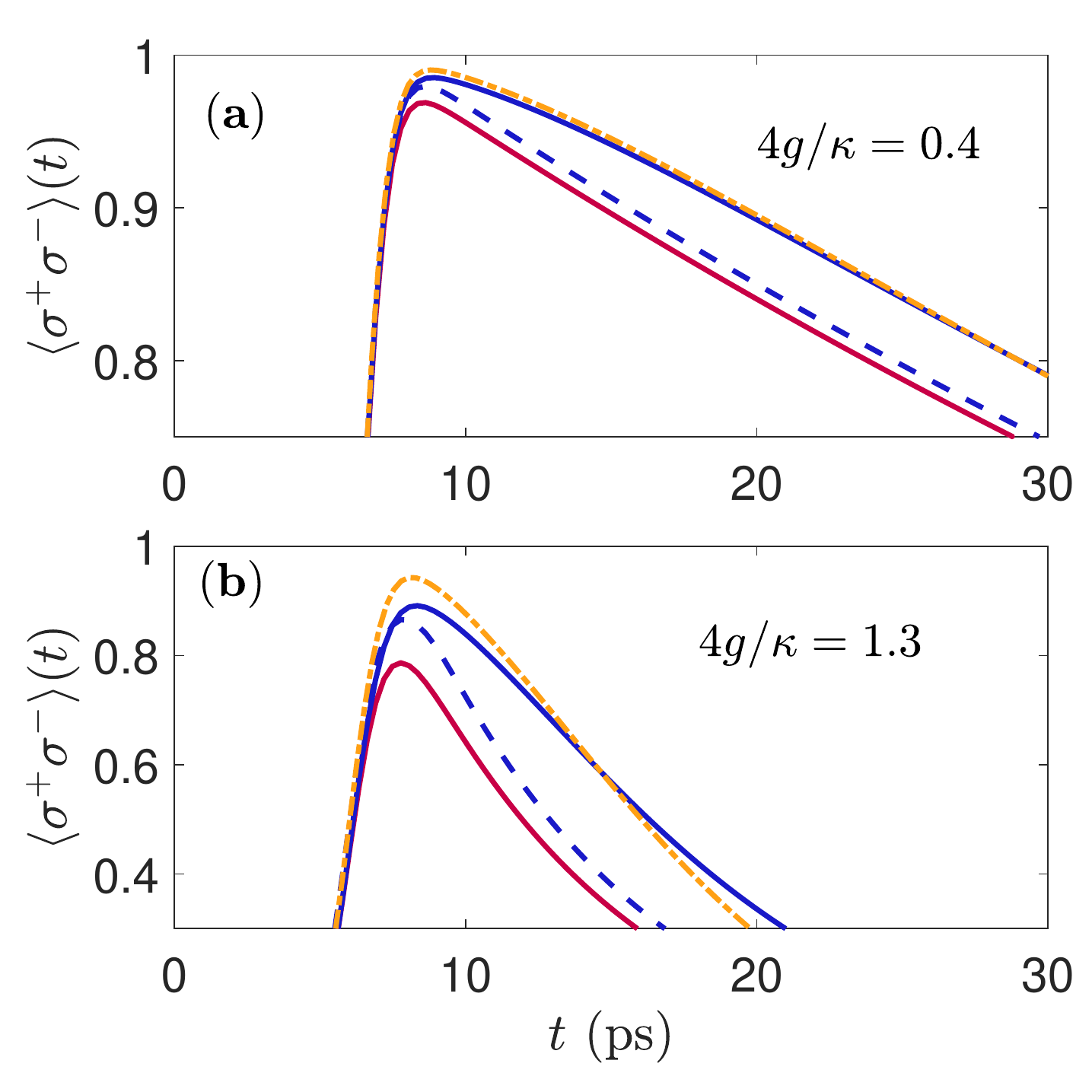}
 \caption{\small Population dynamics of QD-cavity system driven by a $\tau_p = 2 \ \text{ps}$ pulse with the phonon interaction neglected. Here, the cavity decay rate is $\hbar \kappa = 200 \ \mu \text{eV}$, and the QD-cavity coupling constant is (a) $\hbar g = 20 \ \mu\text{eV}$ and (b) $\hbar g = 65 \ \mu eV$. Plotted for comparison is the exact solution (cavity-QED; dash-dotted orange), the system-reservoir model with (dashed blue) and without (solid blue) the additional Markov approximation, and a simple ``bad-cavity" model where $\gamma \rightarrow F_P \gamma$, with $F_P = 4g^2/(\kappa \gamma)$ (solid red).}
\label{fig3} 
\end{figure}

To verify that our approach recovers  the weak-coupling regime ($4g' \ll \kappa$)  cavity-QED model results, where the quantized cavity mode is treated at the system level with creation and destruction operators $a$, $a^\dagger$, we plot in Fig.~\ref{fig3} the exciton population over time for a short pulse, using a series of different models; we have neglected phonons to make the comparison simpler. The additional Markov approximation overestimates the cavity interaction during the pulse, while the system-reservoir result without the additional Markov approximation approaches the exact (cavity-QED) solution in the $4g' \ll \kappa$ regime.

\begin{figure}[!t]
\centering
\includegraphics[width=1\linewidth]{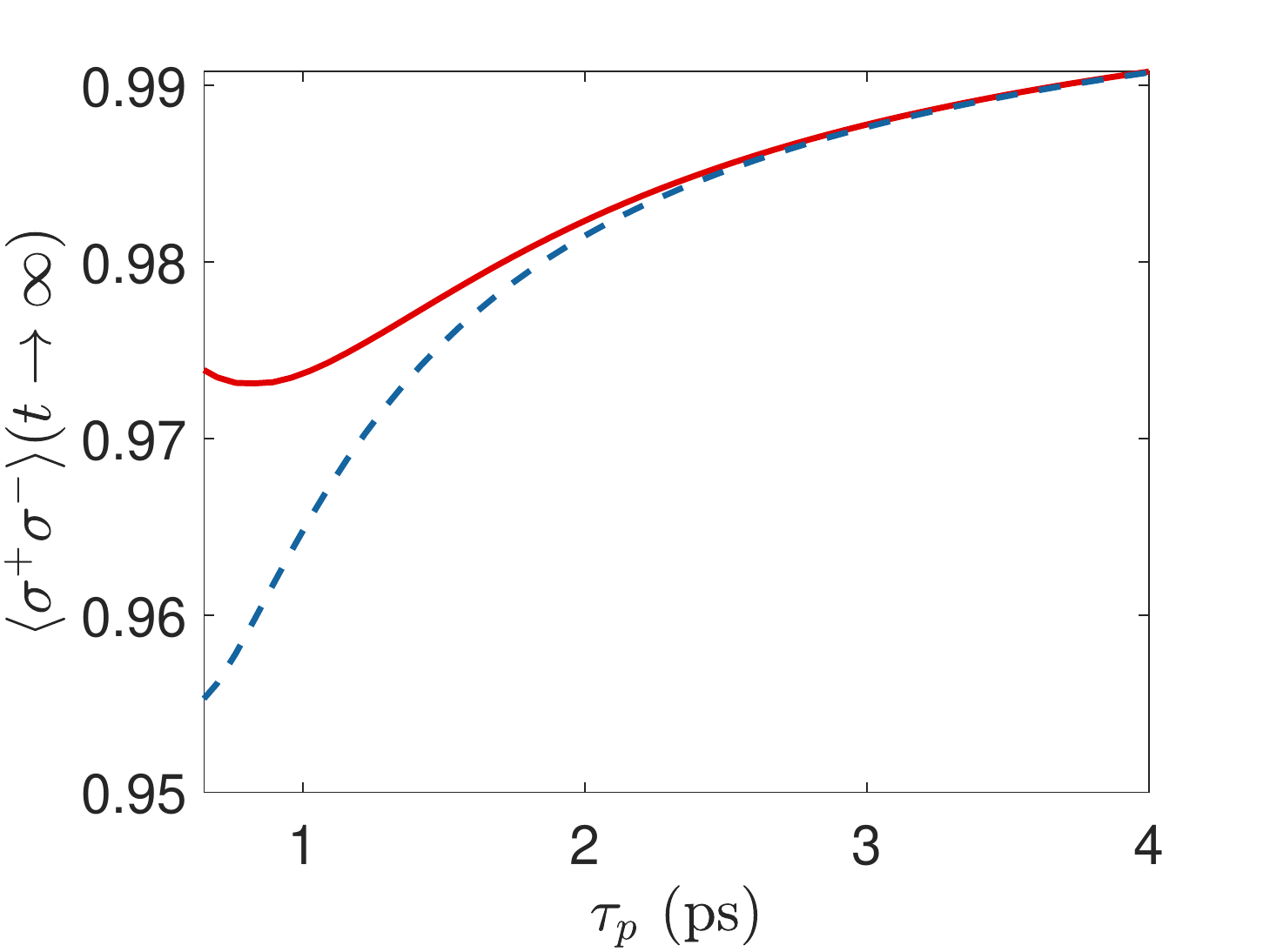}
\caption{\small Inversion efficiency for a driven QD with the additional Markov approximation (blue dashed) and without (red solid). For this calculation, as explained in the text, we have turned off the cavity coupling and spontaneous emission ($\gamma = g = 0$).}
\label{fig4} 
\end{figure}

It is also interesting to assess the effect of retaining the exact unitary operators instead of using the additional Markov approximation on the time-dependent exciton-phonon scattering rates induced by the pulse, as here the phonon bath must be treated as a reservoir in the master equation approach. To isolate the dynamics of the electron-phonon interaction, we ``turn-off" ($g=0$) the cavity coupling and vary the pulse width in Fig.~\ref{fig4}. Note that the lower limit for the pulse width which can be modelled is set by the condition that we remain in the polaron regime at the pulse's maximum amplitude: $\tau_p \gg {\sqrt{\pi(1-\B^4)}}/(\sqrt{2}\omega_b)$, or with our phonon parameters, $\tau_p \gg 0.4 \ \text {ps}$. To be able to define an exciton population steady-state value after the pulse (the population inversion $\lim_{t\rightarrow \infty}\langle \sigma^+\sigma^-\rangle(t)$), we also set the background spontaneous emission $\gamma$ rate to zero, as this merely introduces exponential decay of the exciton population. Retaining the exact interaction picture transformation captures the correct time evolution under the coherent exciton-pulse scattering, and thus the correct coupling with the phonon and photon reservoirs. This figure explains the sudden drop off in emitted photons seen in Fig.~\ref{fig1} with phonons for short pulse widths; this is simply an artifact of the additional Markov approximation overestimating the effect of phonon coupling for short pulse widths. Additionally, the increase in inversion efficiency with increasing pulse width explains the difference in the $N_\text{tot}$ vs. pulse width slopes with and without phonon coupling, which is a consequence of the phonon decoherence rates scaling with the square of the pulse amplitude, while for a given pulse area, the pulse width is inversely proportional to the pulse amplitude. The decrease in system-environment interaction rates for pulse widths comparable or smaller than the reservoir correlation time has a simple physical explanation, namely that the system dynamics become faster than the response time of the environment, or equivalently, the system evolution occurs with frequency components larger than the spectral response functions of the reservoirs. A similar phenomena is predicted at high pump strengths $\Omega$, where the phonon interaction decouples as the system is driven faster (via Rabi oscillations) than the phonon reservoir's correlation time~\cite{vagov07}.

\section{Upper limits to fidelity}\label{sec4}
In this section, we calculate the upper limits to simultaneous efficiency and indistinguishability for a QD-cavity single photon source under resonant pulse excitation, given a QD with a favourable but realistic biexciton binding energy and phonon coupling rates. Before proceeding, it is worth evaluating the strengths and weaknesses of the two models (cavity-QED and system-reservoir) we have outlined above. In general, the cavity-QED approach is more accurate, as it does not make any Markov approximations regarding the QD-cavity interaction, and retains the full quantum correlations between them, while the system-reservoir theory only recovers the exact dynamics in the limit $4g' \ll \kappa$. However, the system-reservoir approach offers other advantages, including physical insight via analytic simplification, ability to model non-cavity photonic environments, and drastically faster computational speed, which is multiple orders of magnitude faster when the additional Markov approximation is not made. Since, as we will show, the optimal regime of QD-cavity single photon source performance does not necessarily satisfy $4g' \ll \kappa$, we shall use the cavity-QED model for this section. To ensure that the correct non-Markovian dynamics are captured with respect to the phonon interaction,  we forgo the additional Markov approximation and calculate exactly the interaction picture operators $\widetilde{X}_m(t-\tau,t)$ by solving the Heisenberg equations of motion for the system Hamiltonian and $\sigma$ and $a$ operators; e.g.,~$\frac{\text{d}}{\text{d}\tau}\sigma^\pm(t-\tau) = -\frac{i}{\hbar}[H'_S(t-\tau),\sigma^\pm(t-\tau)]$ for each value of $t$. This allows us to predict, to high precision, the theoretical upper limits to simultaneous efficiency and indistinguishability for a pi-pulse single photon source. For simplicity, we will assume that the biexciton binding energy is large enough to justify neglecting the biexciton state in the analysis. Note that there is a trade-off between efficiency and indistinguishability as a function of pulse width (see figures~\ref{fig1},~\ref{fig4}); emitted cavity photon number increases with increasing pulse width, due to reduced phonon-induced dephasing rates as well as increasing multiphoton probability, while the indistinguishability decreases due to increasing multiphoton probability, as well as the associated pump-induced and phonon-induced dephasing during the emission process that also occurs when multi-photon emission probability is substantive. We take $\tau_p = 1 \ \text{ps}$ (FWHM $\sim 1.7$ ps) as an optimal pulse width (for a QD with an appropriate biexciton binding energy) and optimize the cavity parameters around this value for the purposes of this section. The main criteria that should be simultaneously satisfied for the optimization of both indistinguishability and efficiency are thus summarized from our findings as follows: the cavity decay rate should be must less than the phonon bath cutoff frequency ($\kappa \ll \omega_b$), to ensure effective cavity filtering of photons from the incoherent phonon sideband~\cite{ilessmith17}. The system should be in the weak-to-intermediate-coupling regime ($2g' \lesssim \kappa$), to avoid unnecessary strong-coupling physics, increased exposure time to background decoherence sources, and increased cavity-induced dephasing. The long-time limit Purcell factor $4g'^2/(\kappa \gamma)$ should be as large as possible to increase the $\beta$-factor (this also requires $\gamma \ll \kappa$), and thus efficiency. To suppress two-photon emission as a result of re-excitation during the pulse, one requires $\kappa\tau_p \ll 1$. To avoid excitation of the biexciton state, $\tau_p E_B/4 \gg 1$. In simple terms, these criteria are often satisfied for weakly-coupled high quality factor cavities (low $\kappa$), with a high Purcell factor, excited by short pulses.

\begin{figure}[ht]
\centering
\includegraphics[width=1\linewidth]{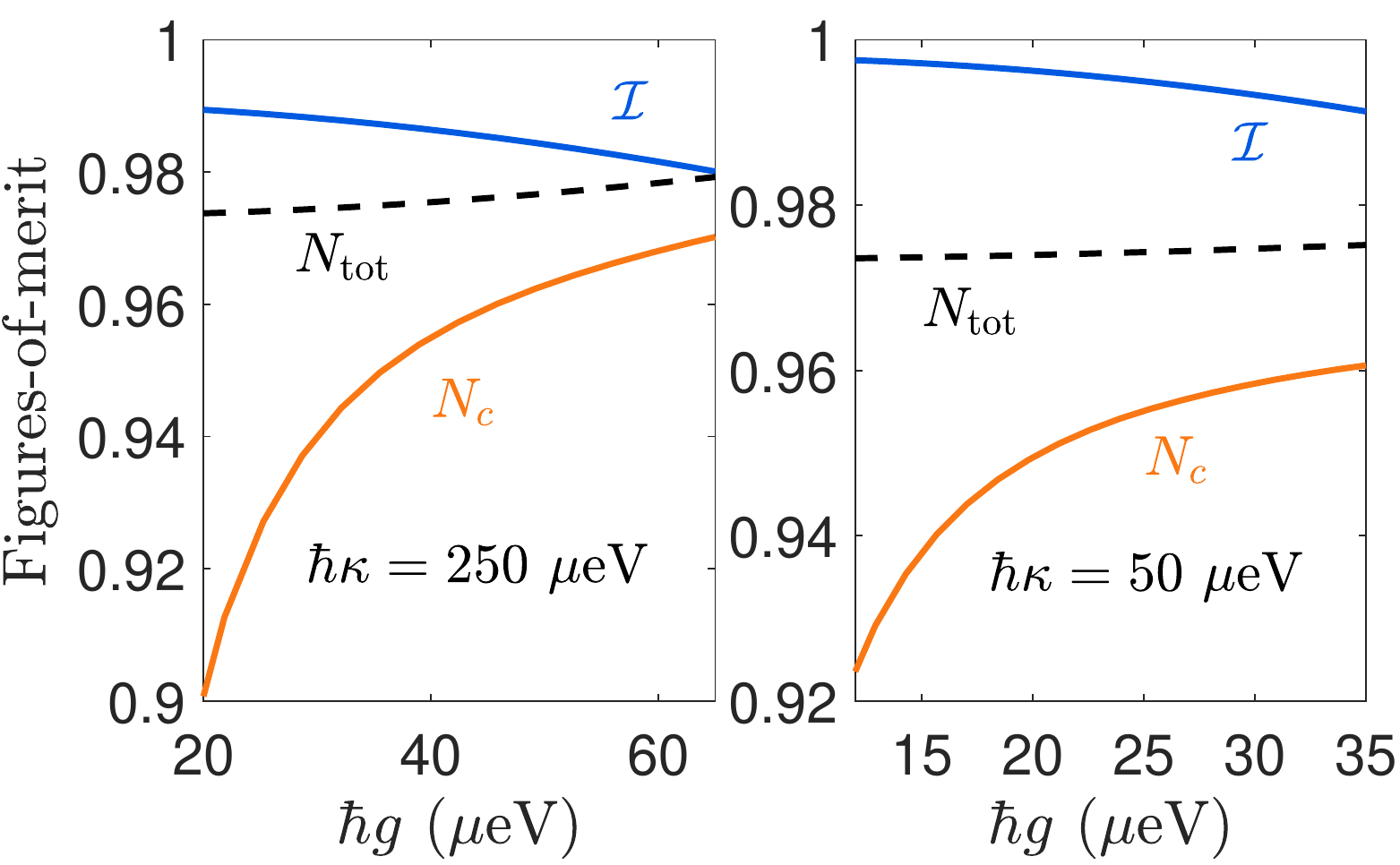}
\caption{\small Figures-of-merit as a function of cavity parameters, using the cavity-QED model without the additional Markov approximation.}
\label{fig5} 
\end{figure}

In Fig.~\ref{fig5}, we plot the single photon sources figures-of-merit for two different cavity decay rates as a function of the coupling strength $g$ to get an estimate of the upper limits to which these parameters can be simultaneously optimized in the presence of pulse-induced decoherence. For reference, by assuming an initially inverted QD, Iles-Smith \emph{et al.}~\cite{ilessmith17} found $\mathcal{I} \approx 99.5\%$ and $N_c = 96\%$ (with the indistinguishability metric converted to the one we use in this paper) at $\hbar g = 30 \ \mu\text{eV}$ and $\hbar \kappa = 120 \ \mu\text{eV}$. For these parameters, we find, with the excitation process included, $\mathcal{I} = 99.3\%$, $N_c = 92.8\%$, where we have used their value of $\hbar\omega_b = 1.025 \ \text{meV}$, and $\hbar\gamma = 1 \ \mu\text{eV}$. As other examples (again using $\hbar\omega_b = 0.9 \ \text{meV}$ and $\hbar\gamma = 0.5 \ \mu\text{ev}$, as with the rest of this paper), we find $\mathcal{I} = 98\%$, $N_c = 97\%$ with  $\hbar g' = 65 \ \mu\text{eV}$, and $\hbar \kappa = 250 \ \mu\text{eV}$, and $\mathcal{I} = 99.6\%$, $N_c = 95\%$ with  $\hbar g' = 20 \ \mu\text{eV}$, and $\hbar \kappa =  50 \ \mu\text{eV}$. We thus conclude that, at least with this method of excitation, the effect of the pulse on the single photon source figures-of-merit is nearly as significant as electron-phonon scattering. Clearly, both phonon scattering and pulse excitation conditions contribute to the limits of single photon source fidelity. It is important to remember that these values are assuming ``best-case" scenario in terms of realistic phonon coupling strengths and biexciton binding energies, and thus constitute an estimate of the upper limit of simultaneous efficiency and indistinguishability. Furthermore, the Purcell factors corresponding to the last two examples given above are quite high: 135.2 and 64 for $\hbar g' = 65 \ \mu\text{eV}$ and $\hbar g' = 20 \ \mu\text{eV}$, respectively, compared with a high Purcell factor seen recently in a QD single photon source at $F_P = 43$~\cite{liu17} (most other sources have had $F_P < 10$). This suggests that the common experimental goal of increasing the Purcell factor in QD single photon sources is well justified.

It should also be noted that the single-photon source efficiency given in this work is at least twice as large as the effective efficiency; to filter the emitted photon from the excitation pulse, the QD is typically excited with a pulse polarized $45\degree$ relative to the QD exciton axes, and a filter orthogonal to this is then applied, filtering out 50\% of the emitted photons. In practice this efficiency loss is larger, as the pulse often couples to an undesired orthogonally polarized cavity mode as well. Thus, for true simultaneous optimization of efficiency and indistinguishability, a different method of source excitation is required. Examples of this include rapid adiabatic passage~\cite{wei14,simon11}, and STIRAP~\cite{gustin17}), although both of these methods require larger pulse lengths, which increases the exposure time to spontaneous emission (and any background pure dephasing), decreasing the single photon figures-of-merit through the processes outlined in this work. One method of excitation which can use short pulses and has the added benefit of strongly suppressing multi-photon emission has been demonstrated recently~\cite{schweickert18,hanschke18}, where the biexciton state is directly pumped via a resonant two-photon excitation pulse, and a cavity resonant with the biexciton-exciton transition extracts on-demand single-photons from the biexciton radiative decay. While this set-up can pose additional challenges (e.g. phonon-absorption-assisted excitation of the exciton during two-photon pumping), for large biexciton binding energies, the exciton can be adiabatically eliminated from the equations of motion modelling the two-photon excitation process of the biexciton state, and thus, much of the analysis of this work applies to this system.

\section{Conclusions}\label{sec5}

To conclude, we have investigated the role that the pulse excitation process plays in QD single-photon sources, including the quantitative influence of two-photon biexciton excitation, multi-photon probability, and phonon-mediated excitation-induced dephasing. Our findings have been expressed in terms of experimentally accessible figures-of-merit, and the models we have presented can be compared with experiment by, e.g., modifying the inversion pulse width (Fig.'s~\ref{fig1},~\ref{fig4}), or the cavity parameters (Fig.~\ref{fig5}). We have also developed a general open-system master equation framework for modelling quantum dots driven by time dependent fields in the presence of acoustic phonon coupling and arbitrary weakly-coupled photonic environments. This framework can model systems driven on timescales comparable to or quicker than the time on which correlations with the environment decay, allowing for non-Markovian effects to be captured. We have elucidated the role these effects play in the excitation of the QD single-photon source, and shown how the cavity and pulse parameters should be optimized for simultaneous maximization of the single-photon indistinguishability and efficiency. 
\section*{aknowledgements}
This work was supported by the Natural Sciences and Engineering Research Council of Canada (NSERC) and Queen's University.
\appendix
\section{Low-temperature phonon-scattering term}\label{appA}
If one is working in the weak phonon coupling regime (low temperatures and/or low phonon coupling constant), and with pulses long enough such that the additional Markov approximation holds, $\mathbb{L}_p$ can be cast in a simpler form by neglecting multi-phonon scattering events by expanding the polaron Green functions to first order in $\phi(\tau)$~\cite{nazir08}. In this case, we have $\Omega(t-\tau) \approx \Omega(t)$, $R(t,\tau) \approx -\frac{1}{2}\Omega(t)\tau$, $G_g(\tau) \approx 0$, and $G_u(\tau) \approx \langle B \rangle^2 \phi(\tau)$. We then find
\begin{align}
\mathbb{L}_p \rho = &\frac{\Gamma_y(t)}{2}\mathcal{L}[\sigma_y]\rho \nonumber \\ +  & \big(\Gamma_u^R(t)[\sigma^+\sigma^-\rho,\sigma_y] + i\Gamma_u^I[\sigma^+\sigma^-,\rho\sigma_y] + \text{H.c.}\big),
\end{align}
where 
\begin{equation}
\Gamma_y(t) = \frac{[\Omega'(t)]^2}{2}\text{Re}\big\{\phi^R(\Omega'(t))\big\},
\end{equation} 
\begin{equation}\Gamma_u^R(t) = \frac{[\Omega'(t)]^2}{2}\text{Im}\big\{\phi^R(\Omega'(t))\big\},
\end{equation}
 and 
\begin{equation}
\Gamma_u^I(t) = \frac{[\Omega'(t)]^2}{2}\text{Im}\big\{\phi^I(\Omega'(t))\big\},
\end{equation}
with $\phi^R(\tau) = \text{Re}\{\phi(\tau)\}$, $\phi^I(\tau) = \text{Im}\{\phi(\tau)\}$, and $\phi^i(\omega) = \int_0^\infty \phi^i(t)e^{i\omega t}dt$. Two of these rates can be simplified analytically: $\Gamma_y(t) = \frac{\pi}{4}J_p\big(\Omega'(t)\big)\coth{\big(\frac{\hbar\Omega'(t)}{2k_BT}\big)}$, and $\Gamma_u^I(t) = -\frac{\pi}{4}J_p\big(\Omega'(t)\big)$. Here it is clear that, within the additional Markov approximation, the exciton adiabatically samples the phonon spectral function directly via the drive amplitude.

\section{Full exciton-cavity scattering terms}\label{appB}
In this appendix, we report $\mathbb{L}_c$ in full generality, without making the approximation that $C_p(\tau) \approx \langle B \rangle ^2$:
\begin{align}
\mathbb{L}_c \rho& = \frac{\Gamma_c(t)}{2}\mathcal{L}[\sigma^-]\rho + \frac{\zeta_c(t)}{2}(\sigma^+\rho\sigma^+ + \sigma^- \rho \sigma^-) \nonumber \\ & -i\tilde{\Omega}(t)(\sigma^+\sigma^-\rho \sigma^+ - \sigma^- \rho\sigma^+\sigma^-)  \nonumber \\ &+ i \!\int_0^\infty \!\! d\tau \text{Im}\big\{\!A_R(\tau)\!\big\}\sin^2{\!\big[R(t,\tau)\big]}(\sigma^+\rho\sigma^+ - \sigma^- \rho \sigma^-)  \nonumber \\ & + \frac{1}{2}\!\int_0^\infty \!\! d\tau \text{Im}\big\{\!A_R(\tau)\!\big\}\sin{\!\big[2R(t,\tau)\big]}([\sigma^+,\sigma_z \rho] + \text{H.c.}) \nonumber \\ &           -i[\frac{\tilde{\Omega}(t)}{2}\sigma_x + \tilde{\Delta}(t)\sigma^+\sigma^-,\rho],
\end{align}
where $A_R(\tau) = C_p(\tau)J_c(\tau) = g'^2\exp[\phi(\tau)-\kappa\tau/2]$, $R(t,\tau) = \frac{1}{2}\int_t^{t-\tau}\Omega'(t')dt'$, and
\begin{equation}
\Gamma_c(t) = 2\int_0^{\infty}d\tau \text{Re}\big\{\!A_R(\tau)\!\big\}\cos^2{\!\big[R(t,\tau)\big]},
\end{equation}
\begin{equation}
\zeta_c(t) = 2\int_0^{\infty}d\tau \text{Re}\big\{\!A_R(\tau)\!\big\}\sin^2{\!\big[R(t,\tau)\big]},
\end{equation}
\begin{equation}
\tilde{\Omega}(t) = -\int_0^{\infty}d\tau \text{Re}\big\{\!A_R(\tau)\!\big\}\sin{\!\big[2R(t,\tau)\big]},
\end{equation}
\begin{equation}
\tilde{\Delta}(t) = \int_0^{\infty}d\tau \text{Im}\big\{\!A_R(\tau)\!\big\}\cos^2{\!\big[R(t,\tau)\big]}.
\end{equation}\\

\bibliography{PRB_arxiv19}
\end{document}